\documentclass[11pt]{article}              
\usepackage{authblk}
\usepackage{amsmath,amsthm}
\usepackage[makeroom]{cancel}
\usepackage{url}
\allowdisplaybreaks
\newtheorem{theorem}{Theorem}
\raggedright                            

\title{\bf Jefimenko Made Easy: Electromagnetic Fields through Retardation}    
\author{Shengchao Alfred Li}            
  
\affil{Amateur scientist in Potomac, Maryland, USA}
\affil{Email: shengchao.li@gmail.com}
\date{Jun. 25, 2023; Updated Oct. 05, 2025}      
\begin{document}                        
\maketitle                              
\begin{abstract}
Oleg D. Jefimenko's electrodynamics textbook is unique in its approaches to deriving the electric and magnetic fields of arbitrary charge and current distributions and of an arbitrarily moving point charge. However, an uncommon form of the inhomogeneous wave equation often poses difficulties for readers right from the beginning. In this paper, we substitute in a commonly used form, making his approaches readily accessible.
\end{abstract}

Oleg D. Jefimenko (1922-2009) was a professor emeritus in the Department of Physics (renamed to the Department of Physics and Astronomy in 2012) at West Virginia University. Persisting into his retirement were his studies in theoretical physics, many of which are controversial. However, his work in electromagnetic retardation, as collected in the first five chapters in his book ``Electromagnetic Retardation and Theory of Relativity: New Chapters in the Classical Theory of Fields, Second Edition" \cite{jefimenko2004electromagnetic}, also found in his earlier textbook \cite{jefimenko1966electricity}, is compatible with other textbooks (thus ``correct") 
\cite{panofsky1962classical} \cite{matveyev1966principles} \cite{podolsky1969fundamentals} \cite{landau1971classical}  \cite{eyges1972classical} \cite{schwartz1972principles} \cite{reitz1979foundations} \cite{barut1980electrodynamics} \cite{becker1982electromagnetic} \cite{lorrain1988electromagnetic} \cite{heald1995classical}  \cite{jackson1998classical} \cite{schwinger1998classical} \cite{griffiths1999introduction} \cite{melia2001electrodynamics} \cite{brau2004modern} \cite{puri2011classical} \cite{steane2012relativity} \cite{franklin2017classical} \cite{davidson2019introduction} \cite{keohane2019introduction} but is unique enough to be worth reading. 

Retardation refers to the phenomenon that it takes time for the effect of a source to propagate to an observer. Starting from Maxwell's Equations and the concept of electromagnetic retardation, without having to rely on the concept of retarded potentials, he derived Jefimenko's Equations, named after him by the popular textbook author David J. Griffiths \cite{griffiths1999introduction}  (also see \cite{heald1995classical}, but see \cite{panofsky1962classical} \cite{mcdonald1997relation} \cite{zangwill2013modern} ), which describe the electric and magnetic fields of an arbitrary charge and current distribution. These equations are usually included in newer textbooks, where they are found by differentiating retarded potentials \cite{lorrain1988electromagnetic} \cite{jackson1998classical} \cite{griffiths1999introduction}  \cite{zangwill2013modern}. From these equations, he derived the fields of several charge and current distributions. He derived the fields of an arbitrarily moving point charge by reducing the size of a moving body of charge, in contrast to the popular but non-straightforward approach of differentiating the Li\'{e}nard-Wiechert potentials (but see \cite{page1922introduction}) in the time domain \cite{panofsky1962classical} \cite{matveyev1966principles} \cite{landau1971classical}   \cite{schwartz1972principles} \cite{reitz1979foundations} \cite{barut1980electrodynamics} \cite{becker1982electromagnetic} \cite{jackson1998classical} \cite{griffiths1999introduction} \cite{melia2001electrodynamics}  \cite{puri2011classical}  \cite{franklin2017classical}  \cite{davidson2019introduction} \cite{keohane2019introduction} or in the frequency domain \cite{podolsky1969fundamentals}.

However, in deriving Jefomenko's Equations, he used an uncommon form of the inhomogeneous wave equations, posing difficulty for readers from the very beginning. In this paper, we replace the uncommon form of the inhomogeneous wave equation in his work with a common one and closely follow Jefimenko's steps. We also add comments in places as needed. Both are aimed at improving understanding.

This common form of inhomogeneous wave equations for the electric field and the magnetic field have been solved without detail \cite{lorrain1988electromagnetic} \cite{mcdonald2018relation} , with the Green function method in the frequency domain \cite{eyges1972classical} \cite{jackson1998classical} \cite{brau2004modern},  or in the time domain \cite{leble2020practical}.

This paper aligns mostly with the first two chapters of Jefimenko's book.

\section{Fields, Sources and Retardation}
In classical electrodynamics, the concept of fields, invented by Faraday, is used in describing electric and magnetic phenomena. 

In a 3-dimensional space, a time-dependent scalar field is expressed as a scalar function $V(x,y,z,t)$ defined at every position $(x,y,z)$ within a region of space and at every time $t$ within a time interval. Similarly, a time-dependent vector field (``vector wave field" by Jefimenko) is expressed as a vector function ${\mathbf V}(x,y,z,t)$. We often omit the coordinates $(x,y,z)$ and $t$ when there is no ambiguity.

The relation between a vector field ${\mathbf V(x,y,z,t)}$ and its source field ${\mathbf Z(x,y,z,t)}$ is often expressed through a differential equation or a set of differential equations. If ${\mathbf Z(x,y,z,t)}$ is given, we can find ${\mathbf V(x,y,z,t)}$ by solving the equation or the set of equations, subject to certain predetermined boundary conditions.


Since it takes time for the effect (e.g., field value at $(x,y,z)$ at time t) to propagate from its cause (e.g., source at $(x^\prime,y^\prime,z^\prime)$ at time $t^\prime$)\footnote{Note that the prime symbol ``$^\prime$" is only used to discriminate between observers and sources. The not-primed and primed coordinates belong to the same coordinate system. Those who have learned special relativity shall not confuse the primed coordinates used here with those used in special relativity conventions.}, $t^\prime$ must be earlier than $t$. If the propagation speed is a fixed speed $c$, we have $t^\prime=t-r/c$ where $r$ is the distance between $(x^\prime,y^\prime,z^\prime)$ and $(x,y,z)$, that is, $r=\sqrt{(x-x^\prime)^2+(y-y^\prime)^2+(z-z^\prime)^2}$. 

We define the retardation of ${\mathbf Z}$ as
\begin{align}
[\mathbf Z]&\stackrel{\operatorname{def}}{=}{\mathbf Z}(x^\prime,y^\prime,z^\prime,t^\prime=t-r/c),\notag
\end{align}
where ``$\operatorname{def}$" stands for ``is defined as"
and $[\ ]$ is called the ``retardation symbol". We call $t^\prime=t-r/c$ retarded time. When there is no ambiguity, we also call the value $r/c$ retardation.





If we utilize the delta function\footnote{It is often called Dirac's delta function, but Heaviside had intensely used it before Dirac.} ($\delta$-function) 
 $\delta(t)$, the retardation of ${\mathbf Z}$ can be expressed as
\begin{align}
 [{\mathbf Z}]&=\int {\mathbf Z}(x^\prime,y^\prime,z^\prime,t^\prime) \delta(t^\prime-(t-r/c)) dt^\prime,\notag
\end{align}
where the effect of the integration is simply to replace $t^\prime$ in ${\mathbf Z}(x^\prime,y^\prime,z^\prime,t^\prime)$ with $t-r/c$.

For future use, we define\footnote{Note that, for the second expression, Jefimenko used the symbol $\left[\frac{\partial{\mathbf Z}}{\partial t}\right]$, which we think may cause confusion. This is because, for the sake of symmetry in notation, we shall either use the pair $[\nabla^\prime {\mathbf Z}]$ and $\left[\frac{\partial{\mathbf Z}}{\partial t^\prime}\right]$, or use the pair $[\nabla{\mathbf Z}]$ and $\left[\frac{\partial{\mathbf Z}}{\partial t}\right]$, but not the pair $[\nabla^\prime {\mathbf Z}]$ and $\left[\frac{\partial{\mathbf Z}}{\partial t}\right]$ as Jefimenko did. Our notation is consistent with the majority textbooks, for example, Jackson's\cite{jackson1998classical}.}
\begin{align}
    [\nabla^\prime {\mathbf Z}]&\stackrel{\operatorname{def}}{=}(\nabla^\prime{\mathbf Z}(x^\prime,y^\prime,z^\prime,t^\prime))|_{t^\prime=t-r/c},\notag\\
    \left[\frac{\partial{\mathbf Z}}{\partial t^\prime}\right] &\stackrel{\operatorname{def}}{=}\left(\frac{\partial{\mathbf Z} (x^\prime,y^\prime,z^\prime,t^\prime)}{\partial t^\prime}\right)\bigg\rvert_{t^\prime=t-r/c},\notag
\end{align}
where the differentiation $\nabla^\prime$ is applied to the primed coordinates, and the symbol ``$|$'' means to plug the subscript into the preceding expression. Similarly, we define
\begin{align}
    [\nabla^\prime \cdot {\mathbf Z}]&\stackrel{\operatorname{def}}{=}(\nabla^\prime\cdot {\mathbf Z}(x^\prime,y^\prime,z^\prime,t^\prime))|_{t^\prime=t-r/c},\notag\\
    [\nabla^\prime \times {\mathbf Z}]&\stackrel{\operatorname{def}}{=}(\nabla^\prime\times{\mathbf Z}(x^\prime,y^\prime,z^\prime,t^\prime))|_{t^\prime=t-r/c}.\notag
\end{align}

\section{Maxwell's Equations and the Inhomogeneous Wave Equations}\label{section2.1} 
\sectionmark{Maxwell's Equations}
Maxwell's Equations in vacuum are,
\begin{align}
    &\nabla\cdot{\mathbf E}=\frac{1}{\epsilon_0}\rho,\label{eqn:maxwell1}\\
    &\nabla\cdot{\mathbf B}=0,\label{eqn:maxwell2}\\
    &\nabla\times{\mathbf E}=-\frac{\partial {\mathbf B}}{\partial t},\label{eqn:maxwell3}\\
    &\nabla\times{\mathbf B}=\mu_0{\mathbf J}+\mu_0\epsilon_0\frac{\partial {\mathbf E}}{\partial t}=\mu_0{\mathbf J}+\frac{1}{c^2}\frac{\partial {\mathbf E}}{\partial t},\label{eqn:maxwell4}
\end{align}
where $\mathbf E$ is the electric field, $\mathbf B$ is the magnetic field, $\rho$ is the charge density, ${\mathbf J}$ is the current density, $\epsilon_0$ is the permittivity of the vacuum,  $\mu_0$ is the permeability of the vacuum, and $\epsilon_0\mu_0=1/c^2$, where $c$ is the speed of light in vacuum.

There is a continuity condition implicitly contained in Maxwell's Equations. We make it explicit by applying the $\nabla\cdot$ operator to both sides of Eq. (\ref{eqn:maxwell4}),
\begin{align}
    \nabla\cdot\nabla\times {\mathbf B}&=\mu_0\nabla\cdot{\mathbf J}+\mu_0\epsilon_0\frac{\partial \nabla\cdot{\mathbf E}}{\partial t}\tag{linearity property}\\
    0&=\nabla\cdot{\mathbf J}+\frac{\partial \rho}{\partial t}.\tag{use Eq. (\ref{eqn:dotcurl}); use Eq. (\ref{eqn:maxwell1})}
\end{align}
Given sources $\rho(x,y,z,t)$ and ${\mathbf J}(x,y,z,t)$, we can obtain ${\mathbf E}(x,y,z,t)$ and ${\mathbf B}(x,y,z,t)$ by solving these equations. In this section, we show that the four equations can be reduced to two inhomogeneous wave equations, one relates ${\mathbf E}$ to its sources, and the other relates ${\mathbf B}$ to its sources.

We apply the $\nabla\times$ operator to both sides of Eq. (\ref{eqn:maxwell3}) and obtain,
\begin{align}
    \nabla\times\nabla\times{\mathbf E}&=\nabla\times\left(-\frac{\partial {\mathbf B}}{\partial t}\right)&\notag\\
    &=-\frac{\partial }{\partial t}\nabla\times{\mathbf B}\tag{linearity property}\\
    &=-\frac{\partial}{\partial t}\left(\mu_0{\mathbf J}+\frac{1}{c^2}\frac{\partial{\mathbf E}}{\partial t}\right)\tag{use Eq. (\ref{eqn:maxwell4})}\\
    &=-\mu_0\frac{\partial{\mathbf J}}{\partial t}-\frac{1}{c^2}\frac{\partial^2{\mathbf E}}{\partial t^2}.\tag{linearity property}
\end{align}
After rearrangement, we have,
\begin{align}
    \nabla\times\nabla\times{\mathbf E}+\frac{1}{c^2}\frac{\partial^2{\mathbf E}}{\partial t^2}&=-\mu_0\frac{\partial {\mathbf J}}{\partial t}.\label{eqn:ihwevEold}
\end{align}
Similarly, we apply the $\nabla\times$ operator to Eq. (\ref{eqn:maxwell4}) and obtain,
\begin{align}
    \nabla\times\nabla\times{\mathbf B}&=\nabla\times \left(\mu_0{\mathbf J}+\frac{1}{c^2}\frac{\partial {\mathbf E}}{\partial t}\right)&\notag\\
    &=\mu_0\nabla\times {\mathbf J}+\frac{1}{c^2}\frac{\partial }{\partial t}\nabla\times{\mathbf E}\tag{linearity property}\\
    &=\mu_0\nabla\times {\mathbf J}+\frac{1}{c^2}\frac{\partial}{\partial t}\left(-\frac{\partial{\mathbf B}}{\partial t}\right)\tag{use Eq. (\ref{eqn:maxwell3})}\\
    &=\mu_0\nabla\times {\mathbf J}-\frac{1}{c^2}\frac{\partial^2{\mathbf B}}{\partial t^2}.\tag{linearity property}\
\end{align}
After rearrangement, we have,
\begin{align}
    \nabla\times\nabla\times{\mathbf B}+\frac{1}{c^2}\frac{\partial^2{\mathbf B}}{\partial t^2}&=\mu_0\nabla\times {\mathbf J}.\label{eqn:ihwevBold}
\end{align}
Note that both Eq. (\ref{eqn:ihwevEold}) and (\ref{eqn:ihwevBold}) are in the same general form,
\begin{align}
	\nabla\times\nabla\times {\mathbf V}(x,y,z,t) +\frac{1}{c^2}\frac{\partial^2 {\mathbf V}(x,y,z,t)}{\partial t^2}={\mathbf K}(x,y,z,t),\notag
\end{align}
where ${\mathbf V}$ is $\mathbf E$ or $\mathbf B$, and ${\mathbf K}$ is $-\mu_0{\partial {\mathbf J}}/{\partial t}$ or $\mu_0\nabla\times {\mathbf J}$, respectively. This is the inhomogeneous wave equation that Jefimenko dealt with.

However, if we plug the following equation
\begin{align}
\nabla\times(\nabla\times{\mathbf E})&=\nabla(\nabla\cdot{\mathbf E})-\nabla^2{\mathbf E}\tag{use Eq. (\ref{eqn:curlcurl})}\\
&=\nabla\left(\frac{1}{\epsilon_0}\rho\right)-\nabla^2{\mathbf E}\tag{use Eq. (\ref{eqn:maxwell1})}\\
&=\frac{1}{\epsilon_0}\nabla\rho-\nabla^2{\mathbf E}\tag{linearity property}
\end{align}
into Eq. (\ref{eqn:ihwevEold}), we get
\begin{align}
    \frac{1}{\epsilon_0}\nabla\rho-\nabla^2{\mathbf E}+\frac{1}{c^2}\frac{\partial^2{\mathbf E}}{\partial t^2}&=-\mu_0\frac{\partial {\mathbf J}}{\partial t},\notag
\end{align}
and after rearrangement, 
\begin{align}
\nabla^2{\mathbf E}-\frac{1}{c^2}\frac{\partial^2{\mathbf E}}{\partial t^2}&=\frac{1}{\epsilon_0}\nabla\rho+\mu_0\frac{\partial{\mathbf J}}{\partial t}.\label{eqn:ihwevE}
\end{align}
Similarly, if we plug the equation
\begin{align}
\nabla\times(\nabla\times{\mathbf B})&=\nabla(\nabla\cdot{\mathbf B})-\nabla^2{\mathbf B}\tag{use Eq. (\ref{eqn:curlcurl})}\\
&=-\nabla^2{\mathbf B}\tag{use Eq. (\ref{eqn:maxwell2})}
\end{align}
into Eq. (\ref{eqn:ihwevBold}), we get
\begin{align}
    -\nabla^2{\mathbf B}+\frac{1}{c^2}\frac{\partial^2{\mathbf B}}{\partial t^2}&=\mu_0\nabla\times {\mathbf J},\notag
\end{align}
and after rearrangement,
\begin{align}    
\nabla^2{\mathbf B}-\frac{1}{c^2}\frac{\partial^2{\mathbf B}}{\partial t^2}&=-\mu_0\nabla\times {\mathbf J}.\label{eqn:ihwevB}
\end{align}
Note that both Eqs. (\ref{eqn:ihwevE}) and (\ref{eqn:ihwevB}) are in the general form, 
\begin{align}
	\nabla^2 {\mathbf V}(x,y,z,t) -\frac{1}{c^2}\frac{\partial^2 {\mathbf V}(x,y,z,t)}{\partial t^2}={\mathbf Z}(x,y,z,t).\label{eqn:ihwev}
\end{align}
where ${\mathbf V}$ is $\mathbf E$ or $\mathbf B$, ${\mathbf Z}$ is ${1}/{\epsilon_0}\nabla\rho+\mu_0{\partial {\mathbf J}}/{\partial t}$ or $-\mu_0\nabla\times {\mathbf J}$, respectively, and $\nabla^2 \mathbf{V}$ is called the Laplacian of $\mathbf{V}$, as defined in Eq. (A.2). This is the inhomogeneous wave equation we usually see in electrodynamics textbooks \cite {johnson1965field} \cite{lorrain1988electromagnetic}  \cite{westgard1995electrodynamics} \cite{brau2004modern} \cite{zangwill2013modern}. In this paper, we shall work with this form instead.

\section{Retarded Integral as a Solution of the Inhomogeneous Wave Equation} \label{RetardedIntegrals}  
\sectionmark{Retarded Integrals}
The vector inhomogeneous wave equation is
\begin{align}
	\nabla^2 {\mathbf V}(x,y,z,t) -\frac{1}{c^2}\frac{\partial^2 {\mathbf V}(x,y,z,t)}{\partial t^2}&={\mathbf Z}(x,y,z,t),\tag{Eq (\ref{eqn:ihwev}) revisited}
\end{align}
where ${\mathbf V}$ and ${\mathbf Z}$ are time-dependent vector fields. We assume ${\mathbf Z}$ is given, and for simplicity, ${\mathbf Z}$ is zero outside a finite region of space. Then ${\mathbf V}$ describes the vector field caused by ${\mathbf Z}$.
\begin{theorem}\label{theorem1}
	The retarded integral 
	\begin{align}
		{\mathbf V}(x,y,z,t)&=-\frac{1}{4\pi}\int \frac{{\mathbf Z}(x^\prime,y^\prime,z^\prime,t^\prime=t-r/c)}{r}dV^\prime\notag\\
		&=-\frac{1}{4\pi}\int \frac{[{\mathbf Z}]}{r}dV^\prime\label{eqn:RetardedIntegral}
	\end{align}
is a solution of the inhomogeneous wave equation Eq. (\ref{eqn:ihwev}), subject to the condition that $[{\mathbf Z}]$ is the cause and ${\mathbf V}$ is the effect, and ${\mathbf V}$ is zero at positions far away from the source, where $r=\sqrt{(x-x^\prime)^2+(y-y^\prime)^2+(z-z^\prime)^2}$, $c$ is the speed of light
and $dV^\prime$ is the infinitesimal volume element at position $(x^\prime, y^\prime, z^\prime)$ of the integration.
\end{theorem}
 
Note that, due to linearity, Eq. (\ref{eqn:ihwev}) can be thought of as having 3 components, each is a scalar inhomogeneous differential equation corresponding to one of the three coordinates, taking the form
\begin{equation}
	\nabla^2 V(x,y,z,t) -\frac{1}{c^2}\frac{\partial^2 V(x,y,z,t)}{\partial t^2}=Z(x,y,z,t),\label{eqn:ihwes}
\end{equation}
where $V$ and $Z$ are scalar fields.

To prove Theorem \ref{theorem1}, due to linearity, we only need to show that each component of Eq. (\ref{eqn:RetardedIntegral}), taking the form
 \begin{align}
{V}(x,y,z,t)=&-\frac{1}{4\pi}\int \frac{[{Z}]}{r}dV^\prime\notag\\
=&-\frac{1}{4\pi}\int \frac{{Z}(x^\prime,y^\prime,z^\prime,t^\prime=t-r/c)}{r}dV^\prime,\label{eqn:RetardedIntegrals}
\end{align}
satisfies the corresponding scalar inhomogeneous wave equation Eq. (\ref{eqn:ihwes}).

This theorem deserves an extensive explanation before we prove it. We address some questions as follows.

1. {\it Why is the wave equation linear?}

Answer: The answer that it is because Maxwell's Equations are linear is valid but somewhat superficial. A deeper answer is that the linearity of the wave equation and of Maxwell's Equations merely reflects the experimental results that the forces or fields produced by the sources are addable. The reason, in turn, may be that the ``stress'' in the vacuum associated with the forces is much below how much the vacuum can handle. 
\qed

2. {\it Why is the solution an integral?}

Answer: This question is tightly related to the first one. Because the wave equation is linear, by definition, we can divide the region occupied by the source ${\mathbf Z}$ into a large number of small volumes. Then, the field value ${\mathbf V}(x,y,z,t)$ becomes the sum of many small field values, each of which is caused by the sources (charge and current) in one of the small volumes. The sum becomes an integral when the sizes of the volumes are made infinitesimal. 
\qed

3. {\it Why does retardation lead to a solution of the wave equation? } 

Answer: 
Retardation and wave propagation are essentially two different names for one phenomenon. If a point source at $(x^\prime,y^\prime,z^\prime)$ emits something\footnote{It helps to imagine that in 1-dimensional space, this ``something" can be stones, bullets; in 2-dimensional space, water surface waves, expanding oil film on a surface; in 3-dimensional space, dust, colored gas, sound waves, etc. Here we have found a common property of moving objects and propagating waves: they move (propagate) in space over time.} at time $t^\prime$ which moves outwards with speed $c$, at time $t$ it will reach position $(x,y,z)$ which lies on the surface of a sphere centered at the source position with diameter $r=c(t-t^\prime)$. 
\qed

4. {\it Why isn't the distance $r$ placed in the retardation symbol $[\ ]$ as $[{\mathbf Z}/r]$}?

Answer: This is because $r$ is only related to the infinitesimal volume elements of the integration and does not depend on retardation, or, the retarded time $t-r/c$. It is a property of space only\footnote{Readers who have learned the Li\'{e}nard-Wiechert potentials shall note that this is radically different from $r$ there.}.
\qed


5. {\it What is special about ${\mathit 1}/r$ so that it plays a role in the solution?} 

Answer: The function $1/r$ is very special because its Laplacian $\nabla^2 (1/r)$ is zero everywhere except at the point $r=0$, and at $r=0$ it is a $\delta$-function, whose volume integration is finite. Proof (following \cite{steane2012relativity}): 
\begin{align}
	\nabla^2\frac{1}{r}&\stackrel{\operatorname{def}}{=}\nabla\cdot\nabla\frac{1}{r}=\nabla\cdot\left(-\frac{1}{r^2}\nabla r\right)\tag{chain rule}\\
    &=\nabla\cdot\left(-\frac{1}{r^2}\nabla (r^2)^\frac{1}{2}\right)\tag{note it is easier to use $\nabla r={\mathbf{i}\frac{\partial r}{\partial x}+\mathbf{j}\frac{\partial r}{\partial y}+\mathbf{k}\frac{\partial r}{\partial z}}$}\\
    &=\nabla\cdot\left(-\frac{1}{r^2}\frac{1}{2}(r^2)^{-\frac{1}{2}}\nabla r^2\right)\tag{chain rule}\\
	&=\nabla\cdot\left(-\frac{1}{r^2}\frac{1}{2r}\nabla r^2\right)=\nabla\cdot\left(-\frac{1}{2r^3}\nabla ({\mathbf r}\cdot{\mathbf r})\right)\tag{a trick: $r^2={\mathbf r}\cdot{\mathbf r}$}\\
	&=\nabla\cdot\left(-\frac{1}{2r^3}2(\nabla{\mathbf r})\cdot{\mathbf r}\right)\tag{use Eq. (\ref{eqn:gradientdotprod})}\\
    &=\nabla\cdot\left(-\frac{1}{r^3}{\mathbf I}{\mathbf r}\right)\tag{$\nabla{\mathbf r}\cdot{\mathbf r}={\mathbf {Ir}}$, see definition in Eq. (\ref{eqn:gradientdotprod})}\\
	&=\nabla\cdot\left(-\frac{\mathbf r}{r^3}\right)\ \mbox{(so far we have proved }\nabla\frac{1}{r}=-\frac{\mathbf r}{r^3}\mbox{ and }\nabla r=\frac{\mathbf r}{r} \mbox{ )}\label{nablar}\\
	&=-\frac{1}{r^3}\nabla\cdot{\mathbf r}-{\mathbf r}\cdot\nabla\frac{1}{r^3}\tag{chain rule, use Eq. (\ref{eqn:divscalarvector})}\\
	&=-\frac{3}{r^3}-{\mathbf r}\cdot\nabla\frac{1}{({\mathbf r}\cdot{\mathbf r})^\frac{3}{2}}\tag{$\nabla\cdot{\mathbf r}=3$; $r=({\mathbf r}\cdot{\mathbf r})^\frac{1}{2}$ from $r^2={\mathbf r}\cdot{\mathbf r}$}\\
	&=-\frac{3}{r^3}+{\mathbf r}\cdot\frac{3}{2}\frac{1}{({\mathbf r}\cdot{\mathbf r})^\frac{5}{2}}\nabla({\mathbf r}\cdot{\mathbf r})\tag{chain rule}\\
	&=-\frac{3}{r^3}+{\mathbf r}\cdot\frac{3}{2}\frac{2{\mathbf r}}{r^5}\tag{for $\nabla({\mathbf r}\cdot{\mathbf r})$, see above or use Eq. (\ref{eqn:gradientdotprod})}\\
	&=-\frac{3}{r^3}+\frac{3r^2}{r^5}\tag{${\mathbf r}\cdot{\mathbf r}=r^2$}\\
        &=\begin{cases}
		0,& \text{if } r \neq 0,\\
		?,& \text{if } r = 0.\mbox{ (can't cancel them out as warned by \cite{steane2012relativity}\footnotemark)}
	\end{cases}\label{cancelzero}
\end{align}
\footnotetext{Steane argued that the reason why the two terms cannot be canceled at $r=0$ is that they are both infinity at $r=0$.}At $r=0$ the value is ill-defined in our calculation, but its integration over the volume of a small ball centered at the origin is well defined. We present here the proof based on Gauss' Law (but see \cite{melia2001electrodynamics} for a different one),
\begin{align}
\int_{\text{ball}}\nabla^2\frac{1}{r}dV&=\int_{\text{ball}}-\nabla\cdot\frac{\mathbf r}{r^3}dV\tag{see remarks in Eq. (\ref{nablar})}\\
&=-\int_{\text{ball}}\nabla\cdot\frac{\hat{\mathbf r}}{r^2}dV\tag{$\hat{\mathbf r}={\mathbf r}/r$}\\
&=-\int_{\text{sphere}}\frac{\hat{\mathbf r}}{r^2}\cdot d{\mathbf S}\tag{to surface integration by Gauss' Law}\\
&=-\frac{1}{r_0^2}\int_{\text{sphere}}\hat{\mathbf r}\cdot d{\mathbf S}=-\frac{1}{r_0^2}\int_{\text{sphere}}dS=-\frac{1}{r_0^2}4\pi r_0^2=-4\pi,\notag
\end{align}
where $r_0$ is the radius of the ball or sphere being integrated with. This result is just the definition of a $\delta$-function,
\begin{align}
	\nabla^2\frac{1}{r}=-4\pi\delta({\mathbf r}).\label{eqn:fourpi}
\end{align}
We will use this equation when we prove the theorem. \qed

6. {\it What really is special about $1/r$?}

Answer: The above mathematical answer to question 5 is valid but not very intuitive. The answer may be in the Taylor expansion of a decreasing function of $r$ that is infinity at $r=0$ and zero at $r=\infty$, with terms depending on $1/r$, $1/r^2$, $1/r^3$, etc. The author of this paper has not yet come up with an intuitive answer. On the other hand, wherever there is a new physics phenomenon, we can dig up or come up with one or more mathematical theories to describe it, with the majority of them, if not all, being approximations. In this case, the theory of $1/r$ is an approximation to the physical reality if the point charge has a non-zero volume. \qed

7. {\it Why is there a coefficient $-\frac{1}{4\pi}$ in the formula?} 

Answer: We inspect Eq. (\ref{eqn:fourpi}) and recognize that a coefficient $-1/(4\pi)$ can normalize $\nabla^2(1/r)$ so that the result, a $\delta$-function, has coefficient one. This, in turn, is related to the fact that the area of the sphere is $4\pi r_0^2$. \qed

8. How can the solution be found mathematically?

Answer: As we have argued previously, the solution of the inhomogeneous scalar wave equation is an integral of the solutions of infinitesimal, or, point sources. 

We derive the solution of a point source by exploiting spherical symmetry\footnote{Jefimenko criticized \cite{becker1982electromagnetic} and \cite{heald1995classical} for their assumption of the source being spherically symmetrical. In hindsight, we can see the authors were talking about a point source thus spherical symmetry assumption should not cause problems. \cite{heald1995classical} was clearer than \cite{becker1982electromagnetic} on this issue but still did not emphasize it enough to prevent misunderstanding.}. In this case, the solution is the Green function \cite{panofsky1962classical} \cite{podolsky1969fundamentals} \cite{barut1980electrodynamics} \cite{jackson1998classical} \cite{melia2001electrodynamics} \cite{franklin2017classical}. It can be derived in the time domain 
\cite{podolsky1969fundamentals} \cite{landau1971classical} \cite{wangsness1986electromagnetic} \cite{pauli2000electrodynamics} \cite{steane2012relativity} \cite{zangwill2013modern} or in the frequency domain \cite{panofsky1962classical} \cite{jackson1998classical} \cite{schwinger1998classical} \cite{pauli2000electrodynamics} \cite{scheck2012classical} . We reproduce the result here in the time domain.

Suppose that, we want to solve the scalar inhomogeneous wave equation with a point source at $(x^\prime,y^\prime,z^\prime)=(0,0,0)$. The equation is $\nabla^2 V-1/c^2\partial^2 V/\partial t=0$ everywhere except at $(0,0,0)$. We first solve the equation for the whole space except for $(0,0,0)$, then we apply the results to $(0,0,0)$.

This problem has spherical symmetry, so we can simplify the equation by using the spherical coordinate system. We assume that since a point source $Z$ is independent of angles, so is $V$. Thus, according to Eq. (\ref{eqn:sphere}) we have
\begin{equation}
    \frac{1}{r}\frac{\partial^2}{\partial r^2}(rV)-\frac{1}{c^2}\frac{\partial^2 V}{\partial t^2}=0.\notag
\end{equation}
Let $u=rV$, we have $V=u/r$. Plug it into the equation, we get,
\begin{align}
    \frac{1}{r}\frac{\partial^2 u}{\partial r^2}-\frac{1}{c^2r}\frac{\partial^2 u}{\partial t^2}=0,\notag
\end{align}
and for places where $r\neq 0$, it reduces to 
\begin{align}
    \frac{\partial^2 u}{\partial r^2}-\frac{1}{c^2}\frac{\partial^2 u}{\partial t^2}=0.\label{uequation}
\end{align}
This two-variable partial differentiation equation is readily solvable with standard methods. We shall omit the derivation here and just write down the general solution of Eq. (\ref{uequation}) as follows,
\begin{align}
    u&=g(t-r/c)+h(t+r/c),\notag
\end{align}
where $g$ and $h$ are arbitrary functions. Thus,
\begin{align}
    V=\frac{g(t-r/c)}{r}+\frac{h(t+r/c)}{r},
\end{align}
where $g(t-r/c)/r$ is called the retarded solution while $h(t+r/c)/r$ is called the advanced solution.


Considering this result, we recognize that, in addition to the retarded particular solution presented in Theorem \ref{theorem1}, 
\begin{align}
		{\mathbf V}_{\mbox{ret}}(x,y,z,t)
		=&-\frac{1}{4\pi}\int \frac{{\mathbf Z}(x^\prime,y^\prime,z^\prime,t^\prime=t-r/c)}{r}dV^\prime,\notag
\end{align}
if we relax the restriction of causality, there is also an advanced particular solution,
\begin{align}
		{\mathbf V}_{\mbox{adv}}(x,y,z,t)
		=&-\frac{1}{4\pi}\int \frac{{\mathbf Z}(x^\prime,y^\prime,z^\prime,t^\prime=t+r/c)}{r}dV^\prime,\label{eqn:advsolution}
\end{align}
where ``ret" stands for ``retarded", and ``adv" ``advanced". Of course, a weighted mixture of the two is also a particular solution. 

The retarded solution is attractive in physics because it is compatible with the idea of causality (but see \cite{wheeler1945interaction} for discussions about the advanced solution).
\qed

9. {\it We learn from the theories that to obtain the general solution of an inhomogeneous partial differential equation, we superpose a particular solution on the general solution of the corresponding homogeneous partial differential equation. How does this reconcile with the fact that we have at least two particular solutions in this case: the retarded integral and the advanced integral?} 


Answer: Of cause, we immediately know from linearity \cite{podolsky1969fundamentals} that the difference
\begin{align}
{\mathbf V}_{\mbox{ret}}(x,y,z,t)-{\mathbf V}_{\mbox{adv}}(x,y,z,t) \label{eqn:homosolution}
\end{align}
is a solution of the homogeneous wave equation,
\begin{align}
	\nabla^2 {\mathbf V}(x,y,z,t) -\frac{1}{c^2}\frac{\partial^2 {\mathbf V}(x,y,z,t)}{\partial t^2}&=0.\label{eqn:homo}
\end{align}
This solution, when added to the advanced integral, gives us the retarded integral.

We shall think about the physical meaning of this solution. It describes a field, infinitesimally small at infinity, whose time-varying ripples, or waves, propagate inward with light speed c, toward where ${\mathbf Z}$ could have been. The waves are not absorbed by ${\mathbf Z}$ because it is missing. They propagate through where ${\mathbf Z}$ could have been, and then propagate outward, toward infinity as if the waves are generated by ${\mathbf Z}$, had ${\mathbf Z}$ existed. 

Of cause, there are other solutions of the homogeneous equation Eq. (\ref{eqn:homo}) if spherical symmetry is not required or the wave does not vanish at infinity. (e.g., a plane wave). These solutions, when superposed with the particular solutions, give us other particular solutions.
\qed

10. {\it Where are the sines and cosines in the ``waves" discussed so far?}

Answer: Waves are not necessarily sinusoidal. In vacuum, waves can be sinusoidal only when the sources, namely, charges or currents, vary sinusoidally. This assertion also applies to media, if we treat induced charges or currents in media as secondary sources. \qed

Now we are ready to prove Theorem \ref{theorem1}. We follow Schwartz's proof \cite{schwartz1972principles} (using the $\delta$-function\footnote{Some authors try to avoid using the $\delta$-function. They divide the whole space into two complementary volumes, including infinitesimally small ball $V_1$ which contains the source position and $V_2$ which did not 
\cite{becker1982electromagnetic} \cite{heald1995classical} \cite{jefimenko2004electromagnetic} \cite{steane2012relativity}, and argued that in $V_1$ the wave equation became the Poisson's equation ($\nabla^2 V \neq 0$). We think this approach is less clear and merely reproduces the work done by the $\delta$-function.}) to show that the solution satisfies the differential equation and the boundary conditions. Although Schwartz proved the corresponding theorem for potentials, the mathematics is similar.

\begin{proof}

We demonstrate that Eq. (\ref{eqn:RetardedIntegrals}) satisfies Eq. (\ref{eqn:ihwes}).
\begin{align}
    \nabla^2 V=&-\frac{1}{4\pi}\int\nabla^2\frac{[Z]}{r}dV^\prime\tag{use Eq. (\ref{eqn:RetardedIntegrals}); linearity property}\\
    =&-\frac{1}{4\pi}\int\nabla\cdot\nabla\frac{[Z]}{r}dV^\prime\tag{definition of $\nabla^2$}\\
    =&-\frac{1}{4\pi}\int \nabla\cdot\left(\frac{\nabla{[Z]}}{r}+{[Z]}\nabla\frac{1}{r}\right)dV^\prime\tag{chain rule}\\
    =&-\frac{1}{4\pi}\int \left(\frac{\nabla^2 [Z]}{r}+\nabla\frac{1}{r}\cdot \nabla [Z]+[Z]\nabla^2 \frac{1}{r}+\nabla [Z]\cdot \nabla \frac{1}{r}\right)dV^\prime\tag{use Eq. (A.4), note that $\nabla[Z]$ and $\nabla(1/r)$ are vectors}\\
    =&-\frac{1}{4\pi}\int \left(\frac{\nabla^2 [Z]}{r}+2\nabla\frac{1}{r}\cdot \nabla [Z]+[Z]\nabla^2 \frac{1}{r}\right)dV^\prime\tag{combine terms}\\
    =&-\frac{1}{4\pi}\int \left(\frac{\nabla^2 [Z]}{r}-2\frac{1}{r^2}\nabla r\cdot\frac{\partial [Z]}{\partial t^\prime}\nabla t^\prime+[Z]\nabla^2 \frac{1}{r}\right)dV^\prime\tag{chain rule; $[Z]$ depends on $(x,y,z)$ only through $t^\prime$}\\
    =&-\frac{1}{4\pi}\int \left(\frac{\nabla^2 [Z]}{r}+2\frac{1}{r^2}\nabla r\cdot\frac{\partial [Z]}{\partial t^\prime}\frac{1}{c}\nabla r+[Z]\nabla^2 \frac{1}{r}\right)dV^\prime\tag{use $t^\prime=t-r/c$}\\
    =&-\frac{1}{4\pi}\int \left(\frac{\nabla^2 [Z]}{r}+\frac{2}{cr^2}\frac{\partial [Z]}{\partial t^\prime}+[Z]\nabla^2 \frac{1}{r}\right)dV^\prime. \notag\\
    &\ \ \ \ \ \ \ \ \ \ \ \ \ \ \ \ \ \ \ \ \ \ \ \ \ \ \ \ \ \ \ \ \ \ \ \ \ \ \ \ \ \ \ \ \ \ \ \mbox{(see remarks in Eq. (\ref{nablar}))}\label{eqnlaplacian}
\end{align}
The first term has $\nabla^2[Z]$ which we process as follows,
\begin{align}
    \nabla^2 [Z]=&\nabla\cdot(\nabla [Z])\notag\\
    =&\nabla\cdot\left(\frac{\partial [Z]}{\partial t^\prime}\nabla t^\prime\right)\tag{$[Z]$ depends on $(x,y,z)$ only through $t^\prime$}\\
    =&\nabla\frac{\partial [Z]}{\partial t^\prime}\cdot\nabla t^\prime+\frac{\partial [Z]}{\partial t^\prime}\nabla\cdot\nabla t^\prime\tag{use Eq. (\ref{eqn:divscalarvector})}\\
    =&\frac{\partial {\nabla [Z]}}{\partial t^\prime}\cdot\nabla t^\prime+\frac{\partial [Z]}{\partial t^\prime}\nabla\cdot\nabla t^\prime\tag{linearity property}\\
    =&\frac{\partial^2 [Z]}{\partial t^{\prime 2}}\nabla t^\prime\cdot\nabla t^\prime+\frac{\partial [Z]}{\partial t^\prime}\nabla\cdot\nabla t^\prime\tag{$[Z]$ depends on $(x,y,z)$ only through $t^\prime$}\\
    =&\frac{\partial^2 [Z]}{\partial t^{\prime 2}}\frac{1}{c^2}(-\nabla r)\cdot(-\nabla r)-\frac{\partial [Z]}{\partial t^\prime}\frac{1}{c}\nabla\cdot\nabla r\ \ \ \ \ \ \ \ \ \ \ \ \ \mbox{(use $t^\prime=t-r/c$)}\notag\\
    =&\frac{1}{c^2}\frac{\partial^2 [Z]}{\partial t^{\prime 2}}-\frac{1}{c}\frac{\partial [Z]}{\partial t^\prime}\nabla \cdot \frac{\mathbf r}{r}\ \ \ \ \ \ \ \ \mbox{(see remarks in Eq. (\ref{nablar}))}\\
    =&\frac{1}{c^2}\frac{\partial^2 [Z]}{\partial t^2}-\frac{1}{c}\frac{\partial [Z]}{\partial t^\prime}\nabla \cdot \frac{\mathbf r}{r} \tag{use $\frac{\partial^2 [Z]}{\partial t^{\prime 2}}=\frac{\partial^2 [Z]}{\partial t^2}$}\\
    =&\frac{1}{c^2}\frac{\partial^2 [Z]}{\partial t^2}-\frac{1}{c}\frac{\partial [Z]}{\partial t^\prime}\left(\frac{1}{r}\nabla\cdot{\mathbf r}+\nabla \frac{1}{r}\cdot{\mathbf r}\right)\tag{use Eq. (\ref{eqn:divscalarvector})}\\
    =&\frac{1}{c^2}\frac{\partial^2 [Z]}{\partial t^2}-\frac{1}{c}\frac{\partial [Z]}{\partial t^\prime}\left(\frac{3}{r}-\frac{1}{r^2}\nabla r\cdot{\mathbf r}\right)\tag{$\nabla\cdot r=3$ and chain rule}\\
    =&\frac{1}{c^2}\frac{\partial^2 [Z]}{\partial t^2}-\frac{1}{c}\frac{\partial [Z]}{\partial t^\prime}\left(\frac{3}{r}-\frac{1}{r}\right)\tag{see remarks in Eq. (\ref{nablar})}\\
    =&\frac{1}{c^2}\frac{\partial^2 [Z]}{\partial t^2}-\frac{2}{cr}\frac{\partial [Z]}{\partial t^\prime}\tag{see next footnote for combining terms}
\end{align}
Plug this result back into Eq. (\ref{eqnlaplacian}) and get
\begin{align}
    \nabla^2 V=&-\frac{1}{4\pi}\int \left(\frac{1}{r}\left(\frac{1}{c^2}\frac{\partial^2 [Z]}{\partial t^2}-\frac{2}{cr}\frac{\partial [Z]}{\partial t^\prime}\right)+\frac{2}{cr^2}\frac{\partial [Z]}{\partial t^\prime}+[Z]\nabla^2 \frac{1}{r}\right)dV^\prime\notag\\
    =&-\frac{1}{4\pi}\int \left(\frac{1}{r}\frac{1}{c^2}\frac{\partial^2 [Z]}{\partial t^2}+[Z]\nabla^2 \frac{1}{r}\right)dV^\prime\tag{cancel out terms\footnotemark}\\
    =&-\frac{1}{4\pi c^2}\frac{\partial^2}{\partial t^2}\int \frac{[Z]}{r}dV^\prime-\frac{1}{4\pi}\int[Z]\nabla^2 \frac{1}{r}dV^\prime\tag{linearity property}\\
    =&\frac{1}{c^2}\frac{\partial^2}{\partial t^2}V-\frac{1}{4\pi}\int[Z]\nabla^2 \frac{1}{r}dV^\prime,\tag{use Eq. (\ref{eqn:RetardedIntegral}) for the first term}
\end{align}
\footnotetext{Why could we cancel out terms here but can not in Eq. (\ref{cancelzero})? A not-so-mathematical, but somehow informative answer might be that although the terms here are also infinity at $(0,0,0)$, they are not as ``large" as those in Eq. (\ref{cancelzero}). After all, if terms in $1/r^3$ integrates to non-zero in an infinitesimal ball, terms in $1/r^2$ and lower orders integrate to $0$.}
and after rearrangement,
\begin{align}
    \nabla^2 V-\frac{1}{c^2}\frac{\partial^2}{\partial t^2}V=&-\frac{1}{4\pi}\int [Z]\nabla^2 \frac{1}{r}dV^\prime\notag\\
    =&\int Z(x^\prime,y^\prime,z^\prime,t^\prime=t-r/c) \delta({\mathbf r}) dV^\prime\tag{use Eq. (\ref{eqn:fourpi})}\\
    =&Z(x,y,z,t)\tag{property of the $\delta$-function; $t^\prime=t$ at $\mathbf{r}=0$}
\end{align}
\end{proof}

\section[Jefimenko's Equations]{Jefimenko's Equations
   \sectionmark{Jefimenko's Equations}}\label{section2.2}
\sectionmark{Jefimenko's Equations}
We apply Theorem \ref{theorem1} directly to the inhomogeneous wave equations Eq. (\ref{eqn:ihwevE}) and Eq. (\ref{eqn:ihwevB}) and obtain
\begin{align}
    {\mathbf E}=&-\frac{1}{4\pi\epsilon_0}\int\frac{\left[\nabla^\prime\rho+\dfrac{1}{c^2}\dfrac{\partial {\mathbf J}}{\partial t^\prime}\right]}{r}dV^\prime,\label{eqn:retardedE}\\
    {\mathbf B}=&\frac{\mu_0}{4\pi}\int\frac{[\nabla^\prime\times{\mathbf J}]}{r}dV^\prime.\label{eqn:retardedB}
\end{align}
These equations relate ${\mathbf E}$ and ${\mathbf B}$ to their causal sources, and are the particular solutions of Maxwell's Equations subject to the boundary conditions that the fields are zero at positions far, far away from the sources, and that causality is assumed. 

By manipulating these two equations, we can derive related formulas,
\begin{align}
    {\mathbf E}&=-\frac{1}{4\pi\epsilon_0}\int\frac{\left[\nabla^\prime\rho+\dfrac{1}{c^2}\dfrac{\partial {\mathbf J}}{\partial t^\prime}\right]}{r}dV^\prime\notag\\
    &=-\frac{1}{4\pi\epsilon_0}\int\frac{\left[\nabla^\prime\rho\right]}{r}dV^\prime
    -\frac{1}{4\pi\epsilon_0}\int\frac{\left[\dfrac{1}{c^2}\dfrac{\partial {\mathbf J}}{\partial t^\prime}\right]}{r}dV^\prime\notag\\
    &=-\frac{1}{4\pi\epsilon_0}\int\nabla\frac{\left[\rho\right]}{r}dV^\prime-\frac{1}{4\pi\epsilon_0}\int\nabla^\prime\frac{\left[\rho\right]}{r}dV^\prime-\frac{1}{4\pi\epsilon_0}\int\frac{\left[\dfrac{1}{c^2}\dfrac{\partial {\mathbf J}}{\partial t^\prime}\right]}{r}dV^\prime\tag{use Eq. (\ref{eqn:nablaZr})}\\
    &=-\frac{1}{4\pi\epsilon_0}\int\nabla\frac{\left[\rho\right]}{r}dV^\prime
    -\frac{1}{4\pi\epsilon_0}\int\frac{\left[\dfrac{1}{c^2}\dfrac{\partial {\mathbf J}}{\partial t^\prime}\right]}{r}dV^\prime\ \ \ \ \ \ \ \ \mbox{(use Eq. (\ref{eqn:intgradient})})\label{eqn:Epotentialpre}\\
    &=\frac{1}{4\pi\epsilon_0}\int\left(\frac{[\rho]}{r^3}+\frac{1}{r^2 c}\left[\frac{\partial \rho}{\partial t^\prime}\right]\right){\mathbf r}d V^\prime-\frac{1}{4\pi\epsilon_0 c^2}\int\frac{1}{r}\left[\dfrac{\partial {\mathbf J}}{\partial t^\prime}\right]dV^\prime,
    \mbox{(use (\ref{eqn:gradientZr}))}\label{eqn:JefimenkoA}
\end{align}
and,
\begin{align}
    {\mathbf B}&=\frac{\mu_0}{4\pi}\int\frac{[\nabla^\prime\times{\mathbf J}]}{r}dV^\prime\notag\\
    &=\frac{\mu_0}{4\pi}\int\nabla\times\frac{[{\mathbf J}]}{r}dV^\prime+\frac{\mu_0}{4\pi}\int\nabla^\prime\times\frac{[{\mathbf J}]}{r}dV^\prime\tag{use Eq. (\ref{nablacrossZr})}\\
    &=\frac{\mu_0}{4\pi}\int\nabla\times\frac{[{\mathbf J}]}{r}dV^\prime\ \ \ \ \ \ \ \ \ \ \ \ \ \ \ \ \ \ \ \ \ \ \ \ \ \ \ \ \ \ \ \ \ \ \ \ \ \mbox{(use Eq. (\ref{eqn:intcurl}))}\label{eqn:Bpotentialpre}\\
    &=\frac{\mu_0}{4\pi}\int\left(\frac{[\mathbf J]}{r^3}+\frac{1}{r^2c}\left[\frac{\partial{\mathbf J}}{\partial t^\prime}\right]\right)\times{\mathbf r}\ dV^\prime\ \ \ \ \ \ \ \ \ \ \ \ \ \ \ \ \mbox{(use Eq. (\ref{eqn:curlZr}))}\label{eqn:JefimenkoB}
\end{align}
Eq. (\ref{eqn:JefimenkoA}) and Eq. (\ref{eqn:JefimenkoB}) are the Jefimenko's Equations. There are other tightly related equations \cite{panofsky1962classical} \cite{mcdonald1997relation}.

\section[Retarded Potentials]{Retarded Potentials
   \sectionmark{Retarded Potentials}}\label{sectionRetardedPotentials}
\sectionmark{Retarded Potentials}
Retarded potentials\footnote{Here we treat potentials as tools that simplify calculations. There are physicists who think potentials are real and have real physical effects (e.g., the Aharonov–Bohm effect). In different gauges, there are different potentials.} can be introduced based on the solutions of ${\mathbf E}$ and ${\mathbf B}$.

As to the magnetic field ${\mathbf B}$, we have
\begin{align}
    {\mathbf B}=&\frac{\mu_0}{4\pi}\int\nabla\times\frac{[{\mathbf J}]}{r}dV^\prime\tag{Eq. (\ref{eqn:Bpotentialpre}) revisited}\\
    =&\nabla\times\left(\frac{\mu_0}{4\pi}\int\frac{[{\mathbf J}]}{r}dV^\prime\right)\tag{linearity property}\\
    \stackrel{\operatorname{def}}{=}&\nabla\times{\mathbf A},\notag
\end{align}
if we define retarded vector potential
\begin{align}
    {\mathbf A}=\frac{\mu_0}{4\pi}\int{\frac{[{\mathbf J}]}{r}}dV^\prime.\notag
\end{align}
As to the electric field ${\mathbf E}$, we have
\begin{align}
{\mathbf E}=&-\frac{1}{4\pi\epsilon_0}\int\nabla\frac{\left[\rho\right]}{r}dV^\prime
    -\frac{1}{4\pi\epsilon_0}\int\frac{\left[\dfrac{1}{c^2}\dfrac{\partial {\mathbf J}}{\partial t^\prime}\right]}{r}dV^\prime\tag{Eq. (\ref{eqn:Epotentialpre}) revisited}\\
    =&-\nabla\left(\frac{1}{4\pi\epsilon_0}\int\frac{\left[\rho\right]}{r}dV^\prime\right)
    -\frac{\partial}{\partial (t-r/c)}\left(\frac{\mu_0}{4\pi}\int\frac{[{\mathbf J}]}{r}dV^\prime\right)\tag{linearity property; $\left[\frac{\partial \mathbf {J}}{\partial t^\prime}\right]=\frac{\partial \mathbf{J}(x^\prime,y^\prime,z^\prime,t^\prime)}{\partial t^\prime}|_{t^\prime=t-r/c}=\frac{\partial \mathbf{J}(x^\prime,y^\prime,z^\prime,t-r/c)}{\partial (t-r/c)}$}\\
    =&-\nabla\left(\frac{1}{4\pi\epsilon_0}\int\frac{\left[\rho\right]}{r}dV^\prime\right)
    -\frac{\partial}{\partial t}\left(\frac{\mu_0}{4\pi}\int\frac{[{\mathbf J}]}{r}dV^\prime\right)\notag\\
    \stackrel{\operatorname{def}}{=}&-\nabla\varphi-\frac{\partial {\mathbf A}}{\partial t},\notag
\end{align}
again if we define retarded scalar potential $\varphi$ and borrow retarded vector potential ${\mathbf A}$ from above such that
\begin{align}
    \varphi=&\frac{1}{4\pi\epsilon_0}\int\frac{\left[\rho\right]}{r}dV^\prime\label{eqn:Epotential}\\
    {\mathbf A}=&\frac{\mu_0}{4\pi}\int\frac{[{\mathbf J}]}{r}dV^\prime\label{eqn:Bpotential}.
\end{align}
We collect the two equations relating fields to potentials as follows,
\begin{align}
    {\mathbf E}=&-\nabla\varphi-\frac{\partial {\mathbf A}}{\partial t},\\
    {\mathbf B}=&\nabla\times{\mathbf A}.
\end{align}

\section[Electromagnetic Induction]{Electromagnetic Induction
   \sectionmark{Electrical Induction}}
\sectionmark{Electrical Induction}
Jefimenko argued that since Eq. (\ref{eqn:JefimenkoA}) and (\ref{eqn:JefimenkoB}) show that ${\mathbf E}$ and ${\mathbf B}$ are generated by their corresponding sources, electric induction (changing electric field induces magnetic field) and magnetic induction (changing magnetic field induces electric field) are just illusions.

This argument sheds some light on electromagnetic phenomena. After all, in electromagnetic waves, the electric field and magnetic field are in phase \cite{griffiths1999introduction}, reaching peaks and troughs together, not alternatively.

He then argued that the physics world has long ignored the ``electrokinetic field" he defined, $-\frac{\mu_0}{4\pi}\int [\partial {\mathbf J}/\partial t^\prime])/r\ dV^\prime$ (the second term of Eq. (\ref{eqn:JefimenkoA})). However, since his results are the same as those of classical electrodynamics, this argument is not very attractive.

\section{The Fields of an Arbitrarily Moving Point Charge}

Jefimenko spent chapters 3-5 of his book on finding electric field ${\mathbf E}$ and magnetic field ${\mathbf B}$ of a few case studies of moving charge distributions. However, current distributions are not included, except for current induced by the moving charges, making the case studies not very general. 

A special case about an arbitrarily moving point charge is interesting because it is related to subatomic charged particles and thus is included in most electrodynamics textbooks.

For this special case, a straightforward approach to obtaining $\mathbf{E}$ and $\mathbf{B}$ used by Jefimenko is to find the field expressions up to low-order approximations of Eqs. (\ref{eqn:retardedE}) and (\ref{eqn:retardedB}), then let the size of the body shrink infinitesimally while keeping the total charge constant. 

The results are the same as those obtained in other textbooks with the Li\'{e}nard-Wiechert potential method, so are the results of such a charge moving in a constant velocity, which is also identical to those obtained from special relativity.

\section{Discussion}
Jefimenko's electrodynamics textbook is a complement to other electrodynamics textbooks. The fact that the results regarding the electric and magnetic fields are the same gives us confidence that they are what Maxwell's Equations mean to lead to, and reinforces our confidence in the notion that Maxwell's Equations, being instantenous, are compatible with the idea of retardation and special relativity.


\appendix
\section{Notations and Identities}
\renewcommand{\theequation}{A.\arabic{equation}}
\setcounter{equation}{0}
In this appendix, we collect definitions and identities used in this paper. 

\subsection*{Definitions} 
We represent a vector in Cartesian coordinates as a row vector $(x,y,z)$, or as column vector $\left(\begin{array}{c}x\\y\\z\end{array}\right)=(x,y,z)^T$, where $^T$ stands for transpose, or as $x{\mathbf i}+y{\mathbf j}+z{\mathbf k}$, or as ${\mathbf i}x+{\mathbf j}y+{\mathbf k}z$, where ${\mathbf i}$, ${\mathbf j}$, and ${\mathbf k}$ are unit vectors pointing to the positive direction of the $x$, $y$, and $z$ axes, respectively, and $x$, $y$, and $z$ are coefficients. Note that we reuse $x,y,z$ symbols here, both for the axes and for the coefficients. Note that unit vectors can be written as ${\mathbf i}=(1,0,0)^T$, ${\mathbf j}=(0,1,0)^T$, and ${\mathbf k}=(0,0,1)^T$, thus,
\begin{align}
    x{\mathbf i}+y{\mathbf j}+z{\mathbf k}=&x\left(\begin{array}{c}1\\0\\0\end{array}\right)+y\left(\begin{array}{c}0\\1\\0\end{array}\right)+k\left(\begin{array}{c}0\\0\\1\end{array}\right)=\left(\begin{array}{c}x\\y\\z\end{array}\right).\notag
\end{align}

We write down the space derivatives of a field in Cartesian coordinates, 
Gradient:
\begin{flalign}
\nabla a = {\mathbf i}\frac{\partial a}{\partial x}+{\mathbf j}\frac{\partial a}{\partial y}+{\mathbf k}\frac{\partial a}{\partial z}.&&\notag
\end{flalign}
Divergence:
\begin{flalign}
\nabla \cdot {\mathbf a} = \frac{\partial a_x}{\partial x}+\frac{\partial a_y}{\partial y}+\frac{\partial a_z}{\partial z}.&&\notag
\end{flalign}
Curl:
\begin{flalign}
\nabla \times {\mathbf a} &= \left|\begin{array}{ccc}{\mathbf i}&{\mathbf j}&{\mathbf k}\\\frac{\partial}{\partial x}&\frac{\partial}{\partial y}&\frac{\partial}{\partial z}\\a_x&a_y&a_z\end{array}\right|&\tag{$|\ |$ stands for determinant}\\
&={\mathbf i}\left(\frac{\partial a_z}{\partial y}-\frac{\partial a_y}{\partial z}\right)+{\mathbf j}\left(\frac{\partial a_x}{\partial z}-\frac{\partial a_z}{\partial x}\right)+{\mathbf k}\left(\frac{\partial a_y}{\partial x}-\frac{\partial a_x}{\partial y}\right).&\notag
\end{flalign}
Laplacian:
\begin{flalign}
\nabla^2 a &= \nabla \cdot {\nabla a} = \frac{\partial^2 a}{\partial x^2}+\frac{\partial^2 a}{\partial y^2}+\frac{\partial^2 a}{\partial z^2}.&\notag
\end{flalign}
Laplacian in spherical coordinates:
\begin{flalign}
    \nabla^2 V &= \frac{1}{r}\frac{\partial^2}{\partial r^2}(rV)+\frac{1}{r^2\sin{\theta}}\frac{\partial}{\partial \theta}\left(\sin \theta \frac{\partial V}{\partial \theta}\right) +\frac{1}{r^2\sin^2\theta}\frac{\partial^2V}{\partial \phi^2},&\label{eqn:sphere}
\end{flalign}
where $\phi$ is the azimuthal angle and $\theta$ is the zenith angle.

Laplacian of a vector:
\begin{flalign}
    \nabla^2{\mathbf a}=&{\mathbf i}\nabla^2 a_x+{\mathbf j}\nabla^2 a_y+{\mathbf k}\nabla^2 a_z&\notag\\
    =&{\mathbf i}\nabla\cdot(\nabla a_x)+{\mathbf j}\nabla\cdot(\nabla a_y)+{\mathbf k}\nabla\cdot(\nabla a_z)&\notag\\
    =&{\mathbf i}\nabla\cdot\left({\mathbf i}\frac{\partial a_x}{\partial x}+{\mathbf j}\frac{\partial a_x}{\partial y}+{\mathbf k}\frac{\partial a_x}{\partial z}\right)
    +{\mathbf j}\nabla\cdot\left({\mathbf i}\frac{\partial a_y}{\partial x}+{\mathbf j}\frac{\partial a_y}{\partial y}+{\mathbf k}\frac{\partial a_y}{\partial z}\right)&\notag\\
    &\ \ \ \ \ \ \ +{\mathbf k}\nabla\cdot\left({\mathbf i}\frac{\partial a_z}{\partial x}+{\mathbf j}\frac{\partial a_z}{\partial y}+{\mathbf k}\frac{\partial a_z}{\partial z}\right)&\notag\\
    =&{\mathbf i}\left(\frac{\partial^2 a_x}{\partial x^2}+\frac{\partial^2 a_x}{\partial y^2}+\frac{\partial^2 a_x}{\partial z^2}\right)+{\mathbf j}\left(\frac{\partial^2 a_y}{\partial x^2}+\frac{\partial^2 a_y}{\partial y^2}+\frac{\partial^2 a_z}{\partial z^2}\right)&\notag\\
    &\ \ \ \ \ \ \ +{\mathbf k}\left(\frac{\partial^2 a_z}{\partial x^2}+\frac{\partial^2 a_z}{\partial y^2}+\frac{\partial^2 a_z}{\partial z^2}\right).&\label{eqn:laplacianvector}
\end{flalign}

\subsection*{Identities}
\begin{flalign}
\nabla({\mathbf a}\cdot{\mathbf b})=\nabla{\mathbf a}\cdot{\mathbf b}+\nabla{\mathbf b}\cdot{\mathbf a}.&& \mbox{(definition of }\nabla {\mathbf a}\mbox{ in proof)\ \ \ \ \ \ \ \ \ \ \ \ }\label{eqn:gradientdotprod}
\end{flalign}
\begin{proof}
\begin{align}
	\nabla({\mathbf a}\cdot{\mathbf b})=&\left({\mathbf i}\frac{\partial }{\partial x}+{\mathbf j}\frac{\partial }{\partial y}+{\mathbf k}\frac{\partial }{\partial z}\right)(a_xb_x+a_yb_y+a_zb_z)\notag\\
	=&{\mathbf i}\left(\frac{\partial a_x}{\partial x}b_x+\frac{\partial a_y}{\partial x}b_y+\frac{\partial a_z}{\partial x}b_z\right)+{\mathbf j}\left(\frac{\partial a_x}{\partial y}b_x+\frac{\partial a_y}{\partial y}b_y+\frac{\partial a_z}{\partial y}b_z\right)\notag\\
	&\ \ \ \ \ \ \ \ \ \ +{\mathbf k}\left(\frac{\partial a_x}{\partial z}b_x+\frac{\partial a_y}{\partial z}b_y+\frac{\partial a_z}{\partial z}b_z\right)\notag\\
    &+{\mathbf i}\left(\frac{\partial b_x}{\partial x}a_x+\frac{\partial b_y}{\partial x}a_y+\frac{\partial b_z}{\partial x}a_z\right)+{\mathbf j}\left(\frac{\partial b_x}{\partial y}a_x+\frac{\partial b_y}{\partial y}a_y+\frac{\partial b_z}{\partial y}a_z\right)\notag\\
    &\ \ \ \ \ \ \ \ \ \ +{\mathbf k}\left(\frac{\partial b_x}{\partial z}a_x+\frac{\partial b_y}{\partial z}a_y+\frac{\partial b_z}{\partial z}a_z\right)\notag\\
	=&\left(\begin{array}{ccc}\frac{\partial a_x}{\partial x}&\frac{\partial a_y}{\partial x}&\frac{\partial a_z}{\partial x}\\\frac{\partial a_x}{\partial y}&\frac{\partial a_y}{\partial y}&\frac{\partial a_z}{\partial y}\\\frac{\partial a_x}{\partial z}&\frac{\partial a_y}{\partial z}&\frac{\partial a_z}{\partial z}\end{array}\right)\left(\begin{array}{c}b_x\\b_y\\b_z\end{array}\right)+\left(\begin{array}{ccc}\frac{\partial b_x}{\partial x}&\frac{\partial b_y}{\partial x}&\frac{\partial b_z}{\partial x}\\\frac{\partial b_x}{\partial y}&\frac{\partial b_y}{\partial y}&\frac{\partial b_z}{\partial y}\\\frac{\partial b_x}{\partial z}&\frac{\partial b_y}{\partial z}&\frac{\partial b_z}{\partial z}\end{array}\right)\left(\begin{array}{c}a_x\\a_y\\a_z\end{array}\right)\notag\\
	=&\left(\begin{array}{ccc}\frac{\partial a_x}{\partial x}&\frac{\partial a_x}{\partial y}&\frac{\partial a_x}{\partial z}\\\frac{\partial a_y}{\partial x}&\frac{\partial a_y}{\partial y}&\frac{\partial a_y}{\partial z}\\\frac{\partial a_z}{\partial x}&\frac{\partial a_z}{\partial y}&\frac{\partial a_z}{\partial z}\end{array}\right)^T\left(\begin{array}{c}b_x\\b_y\\b_z\end{array}\right)+\left(\begin{array}{ccc}\frac{\partial b_x}{\partial x}&\frac{\partial b_x}{\partial y}&\frac{\partial b_x}{\partial z}\\\frac{\partial b_y}{\partial x}&\frac{\partial b_y}{\partial y}&\frac{\partial b_y}{\partial z}\\\frac{\partial b_z}{\partial x}&\frac{\partial b_z}{\partial y}&\frac{\partial b_z}{\partial z}\end{array}\right)^T\left(\begin{array}{c}a_x\\a_y\\a_z\end{array}\right)\notag\\
	\stackrel{\operatorname{def}}{=}&(\nabla{\mathbf a})^T{\mathbf b}+(\nabla{\mathbf b})^T{\mathbf a}\stackrel{\operatorname{def}}{=}\nabla{\mathbf a}\cdot{\mathbf b}+\nabla{\mathbf b}\cdot{\mathbf a}.\notag
\end{align}
\end{proof}
Note that $\nabla {\mathbf a}$ (or $\nabla {\mathbf b}$) is a $3\times 3$ matrix. Its dot product with a vector is defined here as the product of the transpose of the matrix and the vector. This definition is consistent with the dot product of two vectors, i.e., ${\mathbf a}\cdot{\mathbf b}={\mathbf a}^T{\mathbf b}$ (a scalar) where ${\mathbf a}$, ${\mathbf b}$ are $3\times 1$ vectors.

This formula, when used in Electrodynamics, is much better than another one used in many textbooks \cite{jefimenko2004electromagnetic} \cite{schwartz1972principles},
\begin{equation}
	\nabla({\mathbf a}\cdot{\mathbf b})=({\mathbf a}\cdot\nabla){\mathbf b}+({\mathbf b}\cdot\nabla){\mathbf a}+{\mathbf a}\times(\nabla\times{\mathbf b})+{\mathbf b}\times(\nabla\times{\mathbf a}),\notag
\end{equation}
because the former formula exposes the physical intuition of the derivative of a (dot) product of two factors, that the result is the sum of two terms, each is the derivative of one factor multiplied by the other. It is a well-known identity, but its first use in the differentiation of electromagnetic potentials, as far as the author knows, is in Longo's master's thesis \cite{longo2016classical}.
\begin{flalign}
    \nabla\cdot (a{\mathbf b})&=\nabla a \cdot {\mathbf b}+a\nabla \cdot {\mathbf b}.&\label{eqn:divscalarvector}
\end{flalign}
\begin{proof}
\begin{align}
\nabla\cdot (a{\mathbf b})=&\frac{\partial}{\partial x}(a b_x)+\frac{\partial}{\partial y}(ab_y)+\frac{\partial}{\partial z}(ab_z)\notag\\
=&\frac{\partial a}{\partial x}b_x+\frac{\partial a}{\partial y}b_y+\frac{\partial a}{\partial z}b_z\notag+a\frac{\partial b_x}{\partial x}+a\frac{\partial b_y}{\partial y}+a\frac{\partial b_z}{\partial z}\notag\\
=&\left(\mathbf{i}\frac{\partial a}{\partial x}+\mathbf{j}\frac{\partial a}{\partial y}+\mathbf{k}\frac{\partial a}{\partial z}\right)\cdot(\mathbf{i}b_x+\mathbf{j}b_y+\mathbf{k}b_z)+a\nabla \cdot {\mathbf b}\notag\\
=&\nabla a \cdot {\mathbf b}+a\nabla \cdot {\mathbf b}.\notag
\end{align}
\end{proof}

\begin{flalign}
\nabla\cdot\nabla\times{\mathbf a} &=0.&\label{eqn:dotcurl}
\end{flalign}
\begin{proof}
\begin{align}
\nabla\cdot\nabla\times{\mathbf a}&=\nabla \cdot\left({\mathbf i}\left(\frac{\partial a_z}{\partial y}-\frac{\partial a_y}{\partial z}\right)+{\mathbf j}\left(\frac{\partial a_x}{\partial z}-\frac{\partial a_z}{\partial x}\right)+{\mathbf k}\left(\frac{\partial a_y}{\partial x}-\frac{\partial a_x}{\partial y}\right)\right)\notag\\
&=\frac{\partial}{\partial x}\left(\frac{\partial a_z}{\partial y}-\frac{\partial a_y}{\partial z}\right)+\frac{\partial}{\partial y}\left(\frac{\partial a_x}{\partial z}-\frac{\partial a_z}{\partial x}\right)+\frac{\partial}{\partial z}\left(\frac{\partial a_y}{\partial x}-\frac{\partial a_x}{\partial y}\right)\notag\\
&=0.\notag
\end{align}
\end{proof}
\begin{flalign}
\nabla \times (\nabla \times {\mathbf a})=\nabla(\nabla \cdot {\mathbf a})-\nabla^2 {\mathbf a}.&&\label{eqn:curlcurl}
\end{flalign}
\begin{proof}
\begin{align}
    \nabla \times (\nabla \times {\mathbf a})=&\left|\begin{array}{ccc}{\mathbf i}&{\mathbf j}&{\mathbf k}\\\frac{\partial}{\partial x}&\frac{\partial}{\partial y}&\frac{\partial}{\partial z}\\\frac{\partial a_z}{\partial y}-\frac{\partial a_y}{\partial z}&\frac{\partial a_x}{\partial z}-\frac{\partial a_z}{\partial x}&\frac{\partial a_y}{\partial x}-\frac{\partial a_x}{\partial y}\end{array}\right|\notag\\
    =&{\mathbf i}\left(\frac{\partial^2a_y}{\partial x\partial y}-\frac{\partial^2a_x}{\partial y^2}-\frac{\partial^2a_x}{\partial z^2}+\frac{\partial^2a_z}{\partial x\partial z}\right)\notag\\
    &+{\mathbf j}\left(\frac{\partial^2a_z}{\partial y\partial z}-\frac{\partial^2a_y}{\partial z^2}-\frac{\partial^2a_y}{\partial x^2}+\frac{\partial^2a_x}{\partial x\partial y}\right)\notag\\
    &+{\mathbf k}\left(\frac{\partial^2a_x}{\partial x\partial z}-\frac{\partial^2a_z}{\partial x^2}-\frac{\partial^2a_z}{\partial y^2}+\frac{\partial^2a_y}{\partial y\partial z}\right),\notag
\end{align}
\begin{align}
    \nabla(\nabla\cdot {\mathbf a})=&\nabla\left(\frac{\partial a_x}{\partial x}+\frac{\partial a_y}{\partial y}+\frac{\partial a_z}{\partial z}\right)\notag\\
    =&{\mathbf i}\left(\frac{\partial^2 a_x}{\partial x^2}+\frac{\partial^2 a_y}{\partial x\partial y}+\frac{\partial^2 a_z}{\partial x\partial z}\right)\notag\\
    &+{\mathbf j}\left(\frac{\partial^2 a_x}{\partial x\partial y}+\frac{\partial^2 a_y}{\partial y^2}+\frac{\partial^2 a_z}{\partial y\partial z}\right)\notag\\
    &+{\mathbf k}\left(\frac{\partial^2 a_x}{\partial x\partial z}+\frac{\partial^2 a_y}{\partial y\partial z}+\frac{\partial^2 a_z}{\partial z^2}\right),\notag
\end{align}
and compare $\nabla(\nabla\cdot {\mathbf a})-\nabla \times (\nabla \times {\mathbf a})$ to $\nabla^2\mathbf{a}$ (see Eq. (\ref{eqn:laplacianvector})) we know that the identity is proved.
\end{proof}

\subsection*{Manipulations of Retarded Quantities}
In our derivations, we often need to perform mathematical manipulations of expressions involving retarded quantities. Retarded quantities often emerge with the ``source" and are with the primed coordinates. Retarded quantities are often represented with the retardation operator $[\ ]$. 

Recall that we let $\nabla$ denote derivatives with respect to the ordinary coordinates and $\nabla^\prime$ with respect to the primed coordinates. We use $\mathbf{r}$ to represent the vector distance from the source at $(x^\prime,y^\prime,z^\prime)$ to the field at $(x,y,z)$ as $\mathbf{r}=\mathbf{i}(x-x^\prime)+\mathbf{j}(y-y^\prime)+\mathbf{k}(z-z^\prime)$ and $r$ the scalar distance $r=\sqrt{(x-x^\prime)^2+(y-y^\prime)^2+(z-z^\prime)^2}$.

\begin{flalign}
    &\frac{\partial}{\partial x}f({\mathbf r})+\frac{\partial}{\partial x^\prime}f({\mathbf r})=0,&\notag\\
    &\frac{\partial}{\partial x}f( r)+\frac{\partial}{\partial x^\prime}f(r)=0,&\notag\\
    &\frac{\partial}{\partial x}{\mathbf f}({\mathbf r})+\frac{\partial}{\partial x^\prime}{\mathbf f}({\mathbf r})=0,&\notag\\
    &\frac{\partial}{\partial x}{\mathbf f}(r)+\frac{\partial}{\partial x^\prime}{\mathbf f}(r)=0,&\label{eqn:comovexxprime}
\end{flalign}
where $f({\mathbf r})$, $f(r)$, ${\mathbf f}({\mathbf r})$ and ${\mathbf f}(r)$ are any scalar or vector functions of the vector distance ${\mathbf r}$ or the scalar distance $r$ between the field position and the source position. 

The intuition of these equations is that, if the field and its source are related only through the distance vector, moving the observation position is equivalent to moving the source position in the opposite direction. This is the familiar Galilean relativity.

\begin{proof}
It suffices to prove the first equation. 
\begin{flalign}
    \frac{\partial}{\partial x}f({\mathbf r})+\frac{\partial}{\partial x^\prime}f({\mathbf r})
    =&\frac{\partial f({\mathbf r})}{\partial(x-x^\prime)}\frac{\partial (x-x^\prime)}{\partial x}+\frac{\partial f({\mathbf r})}{\partial(x-x^\prime)}\frac{\partial (x-x^\prime)}{\partial x^\prime}\notag\\
    =&\frac{\partial f({\mathbf r})}{\partial(x-x^\prime)}(\frac{\partial (x-x^\prime)}{\partial x}+\frac{\partial (x-x^\prime)}{\partial x^\prime})\notag\\
    =&0.\notag
\end{flalign}
\end{proof}
\begin{flalign}
        \frac{[\partial Z / \partial x^\prime]}{r}=\frac{\partial}{\partial x}\frac{[Z]}{r}+\frac{\partial}{\partial x^\prime}\frac{[Z]}{r},&&\notag\\
        \frac{[\partial Z / \partial y^\prime]}{r}=\frac{\partial}{\partial y}\frac{[Z]}{r}+\frac{\partial}{\partial y^\prime}\frac{[Z]}{r},&&\notag\\
        \frac{[\partial Z / \partial z^\prime]}{r}=\frac{\partial}{\partial z}\frac{[Z]}{r}+\frac{\partial}{\partial z^\prime}\frac{[Z]}{r}.&&\label{eqn:partialZr}
\end{flalign}
\begin{proof}
Because of symmetry, we only need to prove the first equation.
\begin{align}
    &\frac{\partial}{\partial x}\frac{[Z]}{r}+\frac{\partial}{\partial x^\prime}\frac{[Z]}{r}\notag\\
    =&[Z]\frac{\partial}{\partial x}\frac{1}{r}+\frac{1}{r}\frac{\partial}{\partial x} [Z]+[Z]\frac{\partial}{\partial x^\prime}\frac{1}{r}+ \frac{1}{r}\frac{\partial}{\partial x^\prime} [Z]\tag{chain rule}\\
    =&[Z]\left(\frac{\partial}{\partial x}\frac{1}{r}+\frac{\partial}{\partial x^\prime}\frac{1}{r}\right)+\frac{1}{r}\left(\frac{\partial}{\partial x} [Z]+\frac{\partial}{\partial x^\prime} [Z]\right)\tag{combine terms}\\
    =&\frac{1}{r}\left(\frac{\partial}{\partial x}\bigg\rvert_{t^\prime} [Z]+\frac{\partial}{\partial t^\prime} [Z]\frac{\partial t^\prime}{\partial x}\right. +\left.\frac{\partial}{\partial x^\prime}\bigg\rvert_{t^\prime} [Z]+\frac{\partial}{\partial t^\prime} [Z]\frac{\partial t^\prime}{\partial x^\prime}\right)\tag{use Eq. (\ref{eqn:comovexxprime})}\\
    =&\frac{1}{r}\left(\frac{\partial}{\partial t^\prime} [Z](\frac{\partial t^\prime}{\partial x}+\frac{\partial t^\prime}{\partial x^\prime})+\frac{\partial}{\partial x^\prime}\bigg\rvert_{t^\prime} [Z]\right)\tag{$[Z]$ depends on $x$ only through $t^\prime$ }\\
    =&\frac{\partial}{\partial x^\prime}\bigg\rvert_{t^\prime} [Z]\tag{use Eq. (\ref{eqn:comovexxprime}), $t^\prime=t-r/c$ is a function of $r$}\\
    =&\frac{[\partial Z/ \partial x^\prime]}{r},\notag
\end{align}
where $\frac{\partial}{\partial x^\prime}|_{t^\prime}$ means when calculating the derivatives, we do not go into $t^\prime$ with the chain rule.
\end{proof}
We notice that if we replace the factor $1/r$ with any function $f(r)$, the equations still hold.

The intuition of these equations is that if we co-move $x$ and $x^\prime$, $y$ and $y^\prime$, or $z$ and $z^\prime$, the net change on the right side of the equations is caused by the change in $Z$ value due to change of the primed coordinates.

\begin{flalign}
		&\frac{[\nabla^\prime Z]}{r}=\nabla\frac{[Z]}{r}+\nabla^\prime\frac{[Z]}{r},&\label{eqn:nablaZr}\\
		&\frac{[\nabla^\prime\times {\mathbf Z}]}{r}=\nabla\times\frac{[{\mathbf Z}]}{r}+\nabla^\prime\times\frac{[{\mathbf Z}]}{r}.&\label{nablacrossZr}
\end{flalign}
\begin{proof}
We notice that these are simple extensions of Eq. (\ref{eqn:partialZr}). 
\end{proof}
\begin{flalign}
\int \partial/\partial x^\prime &f(x^\prime,y^\prime,z^\prime)dx^\prime=0,&\notag\\
\int \partial/\partial y^\prime &f(x^\prime,y^\prime,z^\prime)dy^\prime=0,&\notag\\
\int \partial/\partial z^\prime &f(x^\prime,y^\prime,z^\prime)dz^\prime=0,&\label{eqn:climbhill}
\end{flalign}
if $f(x^\prime,y^\prime,z^\prime)$ is zero outside of a finite region of space.
\begin{proof}
Because of symmetry, we only need to prove the first equation.
\begin{align}
    \int \partial/\partial x^\prime f(x^\prime,y^\prime,z^\prime)dx^\prime&=\int d_{x^\prime}(f(x^\prime,y^\prime,z^\prime)\notag\\
    &=f(\infty,y^\prime,z^\prime)-f(-\infty,y^\prime,z^\prime)=0.\notag
\end{align}
\end{proof}
The intuition is that if we start from the foot on the left side of a hill, climb to the peak, descend to the right, and reach the foot of the hill on the right side, the net height we cover is zero. 

\begin{flalign}
    &\int \nabla^\prime f(x^\prime,y^\prime,z^\prime)=0,&\label{eqn:intgradient}\\
    &\int \nabla^\prime\times f(x^\prime,y^\prime,z^\prime)=0,&\label{eqn:intcurl}
\end{flalign}
if $f(x^\prime,y^\prime,z^\prime)$ is zero outside of a finite region of space.
\begin{proof}
We notice that these are simple extensions of Eq. (\ref{eqn:climbhill}). 
\end{proof}
This proof is simple because we have restricted $f(x^\prime,y^\prime,z^\prime)$ to be zero outside of a finite region of space. A more general version of this boundary condition requires $f$ to vanish at infinity at a sufficiently fast rate, with respect to the distance $r$. 

\begin{flalign}
    \nabla\frac{[Z]}{r}&=-\frac{\mathbf r}{r^3}[Z]-\frac{\mathbf r}{r^2c}\left[\frac{\partial{Z}}{\partial t^\prime}\right].&\label{eqn:gradientZr}
\end{flalign}
\begin{proof}
\begin{align}
    &\nabla\frac{[Z]}{r}\notag\\
    =&[{Z}]\nabla\frac{1}{r}+\frac{1}{r}\nabla [Z]\tag{chain rule}\\
    =&[{Z}]\left(-\frac{\mathbf r}{r^3}\right)+\frac{1}{r}\frac{\partial [Z]}{\partial t^\prime}\nabla t^\prime\tag{see remarks in Eq. (\ref{nablar})}\\
    =&[{Z}]\left(-\frac{\mathbf r}{r^3}\right)+\frac{1}{r}\left[\frac{\partial{Z}}{\partial t^\prime}\right]\nabla (t-\frac{r}{c})\notag\\
    =&[{Z}]\left(-\frac{\mathbf r}{r^3}\right)-\frac{1}{rc}\left[\frac{\partial{Z}}{\partial t^\prime}\right]\nabla r\notag\\
    =&-\frac{\mathbf r}{r^3}[{Z}]-\frac{\mathbf r}{r^2c}\left[\frac{\partial{Z}}{\partial t^\prime}\right].\tag{see remarks in Eq. (\ref{nablar})}
\end{align}
\end{proof}
\begin{flalign}
    &\nabla\times\frac{[{\mathbf Z}]}{r}=\left(\frac{1}{r^3}[{\mathbf Z}]+\frac{1}{r^2c}\left[\frac{\partial{\mathbf Z}}{\partial t^\prime}\right]\right)\times{\mathbf r}.&\label{eqn:curlZr}
\end{flalign}
\begin{proof}
\begin{align}
    &\nabla\times\frac{[{\mathbf Z}]}{r}\notag\\
    =&\nabla\frac{1}{r}\times[{\mathbf Z}]+\frac{1}{r}\nabla\times[{\mathbf Z}]\tag{chain rule}\\
    =&-\frac{\mathbf r}{r^3}\times[{\mathbf Z}]+
    \frac{1}{r}
    \left|\begin{array}{ccc}{\mathbf i}&{\mathbf j}&{\mathbf k}\\ 
    \frac{\partial}{\partial x}&\frac{\partial}{\partial y}&\frac{\partial}{\partial z}\\
    {[}Z_x]&[Z_y]&[Z_z]\\\end{array}\right|
    \tag{see remarks in Eq. (\ref{nablar})}\\
    =&\frac{1}{r^3}[{\mathbf Z}]\times{\mathbf r}+\frac{1}{r}
    \left|\begin{array}{ccc}{\mathbf i}&{\mathbf j}&{\mathbf k}\\ 
    \frac{\partial t^\prime}{\partial x}&\frac{\partial t^\prime}{\partial y}&\frac{\partial t^\prime}{\partial z}\\
    \frac{\partial [Z_x]}{\partial t^\prime}&\frac{\partial [Z_y]}{\partial t^\prime}&\frac{\partial [Z_z]}{\partial t^\prime}\end{array}\right|\notag\\
    =&\frac{1}{r^3}[{\mathbf Z}]\times{\mathbf r}+\frac{1}{r}
    \left|\begin{array}{ccc}{\mathbf i}&{\mathbf j}&{\mathbf k}\\ 
    -\frac{1}{c}\frac{\partial r}{\partial x}&-\frac{1}{c}\frac{\partial r}{\partial y}&-\frac{1}{c}\frac{\partial r}{\partial z}\\
    \frac{\partial [Z_x]}{\partial t^\prime}&\frac{\partial [Z_y]}{\partial t^\prime}&\frac{\partial [Z_z]}{\partial t^\prime}\end{array}\right|\tag{because $t^\prime=t-r/c$}\\
    =&\frac{1}{r^3}[{\mathbf Z}]\times{\mathbf r}+\frac{1}{r}
    \left|\begin{array}{ccc}{\mathbf i}&{\mathbf j}&{\mathbf k}\\ 
    -\frac{1}{rc}r_x&-\frac{1}{rc}r_y&-\frac{1}{rc}r_z\\
    \frac{\partial [Z_x]}{\partial t^\prime}&\frac{\partial [Z_y]}{\partial t^\prime}&\frac{\partial [Z_z]}{\partial t^\prime}\end{array}\right|\notag\notag\\
    =&\frac{1}{r^3}[{\mathbf Z}]\times{\mathbf r}- \frac{1}{r^2c}{\mathbf r}\times\frac{\partial [{\mathbf Z}]}{\partial t^\prime}\notag\\ 
    =&\frac{1}{r^3}[{\mathbf Z}]\times{\mathbf r}- \frac{1}{r^2c}{\mathbf r}\times\left[\frac{\partial {\mathbf Z}}{\partial t^\prime}\right]\notag\\
    =&\left(\frac{1}{r^3}[{\mathbf Z}]+\frac{1}{r^2c}\left[\frac{\partial {\mathbf Z}}{\partial t^\prime}\right]\right)\times {\mathbf r}.
    \notag
\end{align}
\end{proof}

\bibliographystyle{unsrt}
\bibliography{references}
\end{document}